\documentclass[12pt, journal]{IEEEtran}
\usepackage{epsfig,makeidx,color}
\usepackage{graphicx}
\usepackage{amsbsy}
\usepackage{amsmath}
\usepackage{amssymb}
\usepackage{euscript}
\usepackage{subfigure}
\usepackage{psfrag}

\newcommand {\Define} {\stackrel {\Delta} {=}  }

\renewcommand{\baselinestretch}{1.728}
\begin{document}
\title{Constant-Envelope Multi-User Precoding for Frequency-Selective
  Massive MIMO Systems} \author{\IEEEauthorblockN{Saif Khan Mohammed
    and Erik G. Larsson} \thanks{ Erik G. Larsson is with the
    Communication Systems Division, Dept. of Electrical Eng.\ (ISY),
    Link{\"o}ping University, Link{\"o}ping, Sweden.  Saif Khan
    Mohammed was with the Dept. of Electrical Eng.\ (ISY),
    Link{\"o}ping University.  He is now with the Dept. of Electrical
    Eng., Indian Institute of Technology (I.I.T.), Delhi, India.   This work was supported by VR, SSF and ELLIIT.}}

\onecolumn
\maketitle

\begin{abstract}
  We consider downlink precoding in a frequency-selective multi-user
  Massive MIMO system with highly efficient but non-linear power
  amplifiers at the base station (BS). A low-complexity precoding
  algorithm is proposed, which generates constant-envelope (CE)
  signals at each BS antenna. To achieve a desired per-user
  information rate, the extra total transmit power required under the
  per-antenna CE constraint when compared to the commonly used less
  stringent total average transmit power constraint, is small. 
\end{abstract}
\begin{IEEEkeywords}
Massive MIMO, per-antenna constant-envelope.
\end{IEEEkeywords}
\vspace{-5mm}
\section{Introduction and System Model}\label{sec:SysModel}

Massive MIMO refers to a communication system where a base station
(BS) with a large number of antennas ($N$, a few hundreds)
communicates with several user terminals ($M$, a few tens) on the same
time-frequency resource \cite{SPM-paper}.  There has been recent
interest in massive MIMO systems due to their ability to increase the
spectral and power efficiency even with very low-complexity multi-user
detection and precoding \cite{TM,HNQ}. However, physically building
cost-effective and power-efficient large arrays is a challenge.
Specifically in the downlink, the power amplifiers (PAs) used in the
BS should be highly power-efficient. Since there is a trade-off
between the power-efficiency and linearity of the PA \cite{cripps},
highly power-efficient but non-linear PAs must be used. Since
non-linear PAs could introduce distortion, it is desirable that the
input signal has {\em constant envelope} (CE), that, the signal
transmitted from each BS antenna has a constant amplitude
irrespectively of the channel gains and the information symbols to be
communicated.

With this motivation, in \cite{prev_work} we proposed a CE precoding
algorithm for the frequency-flat multi-user MIMO broadcast channel,
see Section \ref{sec-CEprev} for a brief summary.  In this paper, we
make a non-trivial extension of the CE precoding idea to the case of
frequency-selective channels, and propose a low-complexity per-antenna
CE precoding algorithm, see Section \ref{prop_alg}. Since the channel
has memory we propose to jointly optimize the CE signals transmitted
at consecutive time instances. When $N \gg M$, numerical studies for
the frequency-selective channel with i.i.d.\ Rayleigh fading taps
reveal that, to achieve a desired per-user ergodic information rate,
the proposed CE precoding algorithm needs less than $1.0$ dB extra
total transmit power compared to what is required under the less stringent
and commonly used total average transmit power constraint. We also
observe that even under a stringent per-antenna CE constraint, an
$O(N)$ array gain is achievable, i.e., with every doubling in the
number of BS antennas the total transmit power can be reduced by $3$
dB while maintaining a fixed information rate to each user (assuming
the number of users is kept fixed).

The complex baseband constant envelope signal transmitted from the $i$-th BS antenna at time $t$ is of the form
{
\vspace{-1mm}
\begin{eqnarray}
\label{const_env_eqn}
x_i[t] & = & \sqrt{\frac{P_T}{N}} \, e^{j \theta_i[t]} \,\,\,,\,\,\, i = 1,2,\cdots,N,
\end{eqnarray}
} where $j \Define \sqrt{-1}$, $P_T$ is the total power transmitted
from the $N$ BS antennas and $\theta_i[t] \in [-\pi \,,\, \pi)$ is the
phase of the CE signal transmitted from the $i$-th BS antenna at time
$t$.  The equivalent discrete-time complex baseband channel between
the $i$-th BS antenna and the $k$-th user (having a single-antenna) is
modeled as a finite impulse response filter with an impulse response
of duration $L$ samples and denoted by $(h_{k,i}[0]\,,\, h_{k,i}[1]
\,,\, \cdots \,,\, h_{k,i}[L-1])$.  The signal received at the $k$-th
user ($k=1,2,\cdots,M$) at time $t$ is given by {
\vspace{-1mm}
\begin{eqnarray}
\label{recvk_eqn}
y_k[t] & = & \sqrt{\frac{P_T}{N}} \,\, \sum_{i=1}^N \, \sum_{l=0}^{L-1} \, h_{k,i}[l] e^{j \theta_i[t-l]} \,\,+\,\, w_k[t] \,,\,
\end{eqnarray}
}
where $w_k[t] \sim {\mathcal C}{\mathcal N}(0,\sigma^2)$ is the AWGN at the $k$-th user at time $t$ (AWGN is i.i.d.
across time and across the users).
\section{Philosophy of Constant Envelope Precoding}
\label{sec-CEprev}
Suppose that, at time $t$ we are interested in communicating the information symbol $\sqrt{E_k} u_k[t] \in {\mathcal U}_k \subset {\mathbb C}$ to the $k$-th user. Let ${\mathbb E}[\vert u_k[t] \vert^2] = 1 \,,\, k=1,\cdots,M$. Also, let ${\bf u}[t] = (\sqrt{E_1} u_1[t], \cdots , \sqrt{E_M} u_M[t]) \, \in \, {\mathcal U}_1 \times \cdots \times {\mathcal U}_M$ be the vector of information symbols to be communicated at time $t$. 
In \cite{prev_work}, for frequency-flat channels
we had proposed the idea of choosing the transmit phase angles $\theta_1[t],\theta_2[t],\cdots,\theta_N[t]$ in such a way
that the noise-free signal received at each user is very close to the desired
information symbol,
i.e., $\sqrt{P_T/N} \,\, \sum_{i=1}^N \, h_{k,i}[0] e^{j \theta_i[t]} \, \approx \, \sqrt{P_T} \sqrt{E_k} u_k[t]$ for all $k=1,\cdots,M$.
In \cite{prev_work},
the transmit phase angles corresponding to a desired information symbol vector ${\bf u}[t]$ are given by

{
\vspace{-2mm}
\begin{eqnarray}
\label{NLS_prev_eqn}
 (\theta_1^{u}[t],\cdots,\theta_N^{u}[t])  & \hspace{-3mm}  =  & \hspace{-3mm}  \arg \hspace{-4mm} \min_{_{\hspace{-2mm} \theta_i[t] \in [-\pi,\pi) }} \hspace{-1mm}   \sum_{k=1}^M \hspace{-1mm} {\Big \vert} \frac{ \sum_{i=1}^N h_{k,i}[0] e^{j \theta_i[t]}  } { \sqrt{N}} - \sqrt{E_k} u_k[t] {\Big \vert}^2 \nonumber \\
\end{eqnarray}
}
The idea is to minimize the energy of the difference between the received noise-free signal and the desired information symbol for each user
(this difference was also referred to as multi-user interference (MUI) in \cite{prev_work}).
Further, in \cite{prev_work} a low-complexity iterative algorithm was proposed for solving (\ref{NLS_prev_eqn}) near-optimally when $N$ is
sufficiently large. Here, near-optimality 
refers to achieving a per-user MUI energy level below the AWGN variance.

The main reason behind near-optimality of the precoder in (\ref{NLS_prev_eqn}) when $N \gg M $ is the availability of $(N - M) \gg 1$ excess
degrees of freedom. Let ${\bf H} \in {\mathbb C}^{M \times N}$ be the channel gain matrix with $h_{k,i}[0]$ being its $(k,i)$-th entry. Consider a non-CE precoder like the ZF precoder, where ${\bf x}[t] = {\bf H}^H {\big (} {\bf H} {\bf H}^H {\big )}^{-1} {\bf u}[t]$ is the transmit vector (with appropriate transmit power normalization). Note that adding any vector ${\bf v}[t]$ in the null space of ${\bf H}$ to the transmit vector ${\bf x}[t]$ results in the same received vector (i.e., ${\bf H}{\bf x}[t] = {\bf H} ({\bf x}[t] + {\bf v}[t]) = {\bf u}[t] $). Since the null space dimensionality is very large when $N \gg M$, it is likely that
there exists some ${\bf v}[t]$ such that ${\bf x}[t] + {\bf v}[t]$ is a component-wise CE vector, which then implies that the minimum value of the objective function in (\ref{NLS_prev_eqn}) is close to zero, and hence very little MUI is experienced by each user.

We had analytically shown that for a broad class of frequency-flat channels
(including i.i.d. fading), for a fixed $M$ and fixed symbol energy levels ($E_1, \cdots, E_M$), by having a sufficiently large
$N \gg M$ it is always possible to choose the transmit phase angles so that the received signals at the users are arbitrarily close
to the desired information symbols.
In the next section, inspired by \cite{prev_work} we propose a CE precoding algorithm for frequency-selective
channels.
\section{Proposed CE precoding algorithm}
\subsection{Algorithm}
\label{prop_alg}
For frequency-selective channels, the CE precoder of \cite{prev_work} cannot be applied as it stands, because 
due to channel memory we cannot
find the transmit phase angles for each time instance independently of the transmit phase angles at the previous time instances.
We therefore propose to consider transmission in blocks of $T$ channel uses (time instances), and jointly choose the transmit phase angles
$\theta_i[t] \,,\,i=1,2,\cdots,N\,,\,t=1,\cdots,T$ in such a way that the sum of the energies of the difference between the
desired and the received signals (over all users and over all the $T$ channel uses) is minimized. Block-wise transmission
is also motivated by the fact that real world channels have finite coherence time. More exactly, for a given desired
information symbol vector sequence ${\bf u}[1] \,,\, {\bf u}[2] \,,\, \cdots \,,\,  {\bf u}[T]$, the transmit phase angles are given
by the solution to the optimization problem in (\ref{NLS_joint_eqn})
where $\Theta[t] \Define (\theta_1[t],\theta_2[t],\cdots,\theta_N[t])$ is the vector of transmitted phase angles
at time $t$. 
\begin{figure*}
\begin{eqnarray}
\label{NLS_joint_eqn}
(\Theta^{u}[1],\cdots,\Theta^{u}[T])  &  =  &
\arg \hspace{-5mm} \min_{_{\Theta[t] \in [-\pi,\pi)^N }} 
\sum_{t=1}^T \sum_{k=1}^M {\Bigg \vert} \frac{ \sum_{i=1}^N \, \sum_{l=0}^{L-1} \, h_{k,i}[l] e^{j \theta_i[t-l]}   }{\sqrt{N}} -  \sqrt{E_k} u_k[t]   {\Bigg \vert}^2.
\end{eqnarray}
\vspace{-8mm}
\end{figure*}
\begin{figure*}
\begin{eqnarray}
\label{NLS_joint_eqn1}
(\Theta^{u}[1],\cdots,\Theta^{u}[T]) & = & \arg \hspace{-5mm} \min_{_{\Theta[t] \in [-\pi,\pi)^N } }
\sum_{r=1}^{\lceil{T/\tau}\rceil} I_r \nonumber \\
I_r & \Define &  f(\Theta[(r - 1)\tau - L + 2], \cdots , \Theta[\min( T , r\tau)], {\bf u}[(r - 1)\tau + 1], \cdots, {\bf u}[\min( T , r\tau)]) \nonumber \\
& = & \sum_{t=(r -1)\tau + 1}^{\min( T , r\tau)} \sum_{k=1}^M {\Bigg \vert} \frac{ \sum_{i=1}^N \, \sum_{l=0}^{L-1} \, h_{k,i}[l] e^{j \theta_i[t-l]}   }{\sqrt{N}} -  \sqrt{E_k} u_k[t]   {\Bigg \vert}^2.
\end{eqnarray}
\end{figure*}
We assume that each block of $T$ channel uses is preceded by a uplink training phase used for channel estimation (Time Division Duplexed operation with channel reciprocity is assumed). Since the duration of the training phase is proportional to $L$, we must have $T \gg L $ to avoid loss in overall information rate due to the training overhead.
However, in (\ref{NLS_joint_eqn}) a large $T$ is also associated with more optimization variables,
and therefore a higher complexity.
Hence, in the following we present a method where we split the optimization in (\ref{NLS_joint_eqn}) into several separate
optimization problems with each problem having a much lesser number of optimization variables.
This splitting of (\ref{NLS_joint_eqn}) is not necessarily the optimal way to solve it, but numerical studies confirm its near-optimality (see Section \ref{sec-sim}).

In (\ref{NLS_joint_eqn}) we group the MUI terms of $\tau$ consecutive time instances
into one compound term, as shown in (\ref{NLS_joint_eqn1}) ($I_r$ is the $r$-th such compound term, $r = 1,2 \cdots, \lceil T/\tau \rceil$).
$I_r$ is a function of $\Theta^r \Define ( \Theta[(r-1)\tau + 1]^T , \cdots, \Theta[\min(T,r\tau)]^T)^T$, which is the vector of phase angles to be transmitted at time instances $(r - 1)\tau + 1, \cdots, \min(T,r \tau)$. Note that $I_r$ also depends on the phase angles transmitted during
the last L -1 time instances of the previous compound term $I_{r - 1}$ (i.e., $\Theta[(r -1)\tau - L + 2], \cdots, \Theta[(r -1)\tau]$).

If the time duration $\tau$ is sufficiently larger than the channel memory $L-1$, then $I_r$
depends more strongly on $\Theta^r$ and less strongly on the phase angles transmitted in the last L -1 time instances of $I_{r - 1}$.
This intuition led us to develop a sequential CE precoding algorithm, where we start off with $I_1$ and optimize it
as a function of only the transmit phase angles during the first $\tau$ time instances (i.e., $t=1,2,\cdots,\tau$).
In the $r$-th step of this sequential CE precoding algorithm, we minimize $I_r$ as a function of only $\Theta^r$ 
while keeping the values of $\Theta[(r - 1)\tau - L + 2 ], \cdots, \Theta[(r - 1)\tau ]$
fixed to the value assigned to them after the optimization of $I_{r -1}$. The proposed sequential algorithm stops after the $\lceil T/\tau \rceil$-th step. 

Next, we propose a low-complexity iterative method to minimize $I_r$ as a function of $\Theta^r$
(i.e., the $r$-th step of the sequential CE precoding algorithm).
Let us use the notation $I_r = f_r(\Theta^r)$ for brevity.
For any time instance $t \in [ (r-1)\tau + 1 , \cdots , \min(T, r \tau) ]$, let us define
\begin{equation}
L_r(t)  \Define  \min(L -1 \,,\, \min(T, r  \tau) - t)  \,\,\,,\,\,\,
d_r  \Define  \min(\tau \,,\, T - (r-1)\tau).
\end{equation}
Excess degrees of freedom ($N \gg M$) results in most local minimum of the function $f_r(\cdot)$
having a very small value (compared to the AWGN variance). Through numerical studies we also observe that a simple iterative 
algorithm which optimizes one phase angle at a time while keeping the others fixed, achieves very small values for the objective function $f_r(\cdot)$.
Each iteration of  the proposed iterative method is in turn split into $N d_r$ sub-iterations.
In the $(n,q)$-th sub-iteration of any given iteration ($q \in [1 , \cdots, d_r]$, $n \in [1, \cdots, N]$),
we minimize the current value of $I_r$ as a function of only the phase angle
to be transmitted from the $n$-th BS antenna at time $(r-1)\tau + q$ while keeping the other $N d_r - 1$ phase angles
fixed to their value from the previous sub-iteration. The phase angle $\theta_n[(r-1)\tau + q]$ is then updated to the optimal
value resulting from this single variable optimization and the algorithm moves
to the next sub-iteration. In the next sub-iteration, we repeat the same for the phase angle
to be transmitted from the $n+1$-th BS antenna at time $(r-1)\tau + q$. For the special case when $n = N$, the next sub-iteration minimizes $I_r$ as a function of only the phase angle
to be transmitted from the first BS antenna at time $(r-1)\tau + q +1$, i.e., $\theta_1[(r-1)\tau + q + 1]$.
One full iteration is completed once we have finished the sequence of one-dimensional optimizations for all the $N d_r$ transmit
phase angles, after which we start over with the first sub-iteration $(1,1)$ of the next iteration.

The single variable optimization in the $(n,q)$-th sub-iteration is given by (\ref{sing_var_opt}).
At the end of the $(n,q)$-th sub-iteration the value of $\theta_n[(r -1)\tau + q]$ is modified to $\theta_n^\prime[(r -1)\tau + q]$ 
(see (\ref{sing_var_opt})), whereas all the other $N d_r -1$ angles remain unchanged.
Note that the term $S_{n,t_1}(k,t)$ in (\ref{sing_var_opt}) does not depend on $\theta_n[(r -1)\tau + q]$.
Further, in the $(n,q)$-th sub-iteration
\begin{figure*}
\begin{eqnarray}
\label{sing_var_opt}
\theta_n^\prime[(r-1)\tau + q] & = & \arg \min_{\phi \, \in \, [-\pi \,,\, \pi)} f_r{\Big (}\theta_1[(r-1)\tau + 1], \cdots, \theta_N[(r-1)\tau +1], \cdots, \theta_1[(r-1)\tau + q] , \cdots \nonumber \\
& & \hspace{28mm} , \theta_{n -1}[(r-1)\tau + q], \, \phi \, , \theta_{n+1}[(r-1)\tau + q] , \cdots   \nonumber \\
& & \hspace{28mm}  , \theta_N[(r-1)\tau + q], \cdots , \theta_1[(r-1)\tau + d_r], \cdots, \theta_N[(r-1)\tau + d_r]{\Big )} \nonumber \\
& = & \pi \, + \, \arg {\Big \{}  \sum_{t=t_1}^{t_1 + L_r(t_1)} \, \sum_{k=1}^M \, h_{k,n}^*[t - t_1] S_{n,t_1}(k,t) {\Big \}} \,\,\,\,,\,\,\,
\mbox{\small{where}}  \,\,\,\,\,\,
t_1  \, \Define \,  (r -1)\tau + q \nonumber \\
S_{n,t_1}(k,t) & \Define & \frac{ \sum_{i=1}^N \, \sum_{l=0 \,,\, (i,l) \ne (n,(t - t_1))}^{L-1} \, h_{k,i}[l] e^{j \theta_i[t-l]}   }{\sqrt{N}} -  \sqrt{E_k} u_k[t]
\end{eqnarray}
\end{figure*}
{
\vspace{-3mm}
\begin{eqnarray}
\label{sing_var_opt1}
S_{n,t_1}(k,t) & = & S(k,t) \, - \, \frac{h_{k,n}[t - t_1] e^{j \theta_n[(r-1)\tau + q]}}{\sqrt{N}} 
\end{eqnarray}
}
{
\vspace{-10mm}
\begin{eqnarray}
\label{sing_var_opt2}
S(k,t)  & \Define & \frac{ \sum_{i=1}^N \, \sum_{l=0}^{L-1} \, h_{k,i}[l] e^{j \theta_i[t-l]}   }{\sqrt{N}} -  \sqrt{E_k} u_k[t] \nonumber \\
& & \hspace{-18mm} k=1,\cdots,M\,,\,t=(r-1)\tau +1,\cdots,(r-1)\tau + d_r
\end{eqnarray}
}
From (\ref{sing_var_opt}), (\ref{sing_var_opt1}) and (\ref{sing_var_opt2}), it is clear that for computing the optimal $\theta_n^\prime[(r -1)\tau + q]$
we only need to know  $S(k,t)\,,\,k=1,\cdots,M\,,\,t=t_1,\cdots,t_1+L_r(t_1)$ ($t_1 \Define (r - 1)\tau + q$).
In fact, for each sub-iteration we need not compute $S(k,t)$ as a double sum over $(i,l)$ (as defined in (\ref{sing_var_opt2})).
We only need to compute $S(k,t)$ for all $k=1,\cdots,M\,,\,t=(r-1)\tau +1,\cdots,(r-1)\tau + d_r$
only once during the start of the $(1,1)$-th subiteration of the first iteration.
Thereafter, since we only update one phase angle per sub-iteration, the new updated value of $S(k,t)$ at the end
of the $(n,q)$-th sub-iteration is given by
\begin{eqnarray}
S^\prime(k,t) & = & S(k,t) \, + \, \frac{h_{k,n}[t - t_1]}{\sqrt{N}} {\Big (} e^{j \theta_n^\prime[t_1]} \, - \, e^{j \theta_n[t_1]}  {\Big )}  \nonumber \\
& & \hspace{-10mm}  t=t_1, \cdots, t_1 + L_r(t_1) \,\,,\,\,k=1,\cdots,M.
\end{eqnarray}
The value of $S(k,t)$ for the remaining time instances is not changed.
For $\tau \geq L$ the per-iteration complexity of computing all the $NT$ phase angles is $O(NMLT)$, i.e., a per-channel-use complexity of $O(NML)$. Therefore the complexity increases only linearly in the length of the channel impulse response.
Numerically, it has been observed that very few iterations ($\leq 5$) are required to converge to small values of $f_r(\cdot)$.
\subsection{Performance Analysis}
For a given set of information symbol vectors ${\bf u}[t] \,,\,t=1,\cdots,T$,
let the output phase angles of the proposed CE precoding algorithm be denoted by
${\Tilde \Theta}^u[1],{\Tilde \Theta}^u[2],\cdots,{\Tilde \Theta}^u[T]$.
The phase angle to be transmitted from the $n$-th antenna at time $t$ is denoted by ${\Tilde \theta}_n^u[t]$.
The signal received at the $k$-th receiver at time $t$ is then given by
\begin{eqnarray}
\label{recvk_eqn_tilde}
y_k[t]
& = & \sqrt{P_T} \sqrt{E_k} \, u_k[t] \,+\, \sqrt{P_T} I_k^u[t] \, + \, w_k[t] \nonumber \\
 I_k^u[t] & \Define & {\Big (} \frac{ \sum_{i=1}^N \, \sum_{l=0}^{L-1} \, h_{k,i}[l] e^{j {\Tilde \theta}^u_i[t-l]} }{\sqrt{N}} \, - \, \sqrt{E_k}  u_k[t]  {\Big )}
\end{eqnarray}
From (\ref{recvk_eqn_tilde}) we observe that $I_k^u[t]$ essentially behaves like MUI.
We also define the following vector notation for later use, ${\bf y}_k  \Define (y_k[1], \cdots, y_k[T])^T$,
${\bf u}_k \Define (\sqrt{E_k}u_k[1], \cdots,\sqrt{E_k} u_k[T])^T$, ${\bf I}_k^u \Define (I_k^u[1], \cdots, I_k^u[T])^T$ and ${\bf w}_k \Define (w_k[1], \cdots, w_k[T])^T$ .
From (\ref{recvk_eqn_tilde}) it follows that
\begin{eqnarray}
\label{def_zk}
{\bf z}_k \Define {\bf u}_k \,  -  \, {\bf y}_k/\sqrt{P_T}  \, = \,  - {\bf I}_k^u  \, - \, {\bf w}_k/\sqrt{P_T} 
\end{eqnarray}
Let ${\bf H} = \{ h_{k,i}[l] \} \in {\mathbb C}^{N \times M \times L}$ be the collection of all the channel impulse responses between the $N$ BS antennas and the $M$ users. For a given ${\bf H}$, an achievable rate for the $k$-th user is then given by the mutual information
$I({\bf y}_k \, ; \, {\bf u}_k \,\, | \,\, {\bf H} ) / T $ \cite{wolfowitz},\cite{CTbook}.
For any arbitrary input distribution on ${\bf u}_k  \,,\,k=1,\cdots,M$,
it is in general difficult to compute this mutual information in closed form and therefore in the following we derive a lower bound
assuming $\sqrt{E_k} u_k[t] \,,\, t=1,\cdots, T$ to be i.i.d. complex Gaussian
with mean $0$ and variance $E_k$. 
\begin{eqnarray}
\label{inf_rate_bnd}
I({\bf y}_k \, ; \, {\bf u}_k \,\, | \,\, {\bf H} )  & = & h ({\bf u}_k) \, - \, h ({\bf u}_k \, | \, {\bf y}_k \,,\, {\bf H})  \nonumber \\
&  = &   h ({\bf u}_k) \, - \, h ({\bf u}_k \, - \, {\bf y}_k/\sqrt{P_T} \, | \, {\bf y}_k \,,\, {\bf H})  \nonumber \\
&  = &     T \log_2(\pi e E_k)  \, - \, h( {\bf z}_k \, | \,  {\bf y}_k \,,\, {\bf H})  \nonumber \\
&     \geq &  T \log_2(\pi e E_k)  \, - \, h( {\bf z}_k \, | \,   {\bf H}) 
\end{eqnarray}
where ${\bf z}_k$ is defined in (\ref{def_zk}), $h(\cdot)$ denotes the differential entropy function, and the last inequality follows from the fact that conditioning reduces entropy \cite{CTbook}.
Further, for any random complex vector ${\bf z}_k \in {\mathbb C}^T$ with autocorrelation matrix ${\bf R}_z \Define {\mathbb E}[ {\bf z}_k {\bf z}_k^H]$ , $h({\bf z}_k)$ is maximized when ${\bf z}_k$ is a zero mean proper complex Gaussian distributed vector and therefore
$h({\bf z}_k \, | \, {\bf H})  \leq \log_2 ( (\pi \, e)^T \, \vert {\bf R}_z \vert)$ \cite{proper_complex} (here $\vert {\bf R}_z \vert$ is the determinant of ${\bf R}_z$).
From (\ref{def_zk}) it follows that ${\bf R}_z \, = \, {\mathbb E} [  {\bf I}_k^u \, {\bf I}_k^{u^H} \, | \, {\bf H} ] \, + \, (\sigma^2/P_T) {\bf I}$, where the expectation is over ${\bf u}_1, \cdots, {\bf u}_M$. 
Using this upper bound (on $h( {\bf z}_k \, | \,   {\bf H})$) in (\ref{inf_rate_bnd}) we get an achievable rate for the $k$-th user (the lower bound $R_k({\bf H},{\bf E},P_T/\sigma^2)$ in
(\ref{rk_eqn})). ${\bf E} \Define (E_1,E_2,\cdots,E_M)^T$ is the vector of the average information symbol energies of the $M$ users.
\begin{figure*}
\begin{eqnarray}
\label{rk_eqn}
\frac{I({\bf y}_k \, ; \, {\bf u}_k \,\, | \,\,  {\bf H} )}{T } & \hspace{-2mm} \geq & \hspace{-2mm} \max {\Bigg ( } 0 \,,\, 
\log_2(E_k)    \,  - \,  \frac{\log_2{\Big (}{\Big \vert}  {\mathbb E} [  {\bf I}_k^u \, {\bf I}_k^{u^H} \, | \, {\bf H} ] \, + \, (\sigma^2/P_T) {\bf I}      {\Big \vert} {\Big )}}{T }  {\Bigg ) }  \, \Define \, R_k({\bf H},{\bf E},\frac{P_T}{\sigma^2})
\end{eqnarray}
\end{figure*}
The ergodic mutual information lower bound for the $k$-th user is then given by ${\mathbb E}[ R_k({\bf H},{\bf E},P_T/\sigma^2) ]$ (expectation is over ${\bf H}$).
\section{Numerical results and discussion}
\label{sec-sim}
We consider a Rayleigh fading frequency selective channel with a uniform average power delay profile, i.e., ${\mathbb E}[\vert h_{k,i}[l] \vert^2] = 1/L \,,\,l=0,1,\cdots,(L-1)$.
The channel gains $h_{k,i}[l]$ are i.i.d. circularly symmetric complex Gaussian (mean $0$, variance $1/L$).
We consider the ergodic sum rate $\sum_{k=1}^M {\mathbb E}[ R_k({\bf H},{\bf E},P_T/\sigma^2) ]$ as the measure of performance.
In order to further optimize the ergodic sum rate, we would need to maximize it as a 
function of $E_1,\cdots,E_M$. This is however a difficult task. Since the users have identical channel
statistics, it is likely that the optimal ${\bf E}$ vector is such that $E_1=E_2= \cdots=E_M=E^*$.
Using numerical methods, for a given $P_T/\sigma^2$ we therefore find the optimal $E^*$ which results in the largest ergodic sum rate.
With $E_1=E_2= \cdots=E_M=E^*$, we also numerically observe that ${\mathbb E}[ R_1({\bf H},{\bf E},P_T/\sigma^2) ] = \cdots = {\mathbb E}[ R_M({\bf H},{\bf E},P_T/\sigma^2) ]$, i.e., all users have the same ergodic rate.
Let us denote this per-user ergodic achievable rate by $R(N,M,P_T/\sigma^2)$.

\begin{figure}[t]
\vspace{-1mm}
\begin{center}
\epsfig{file=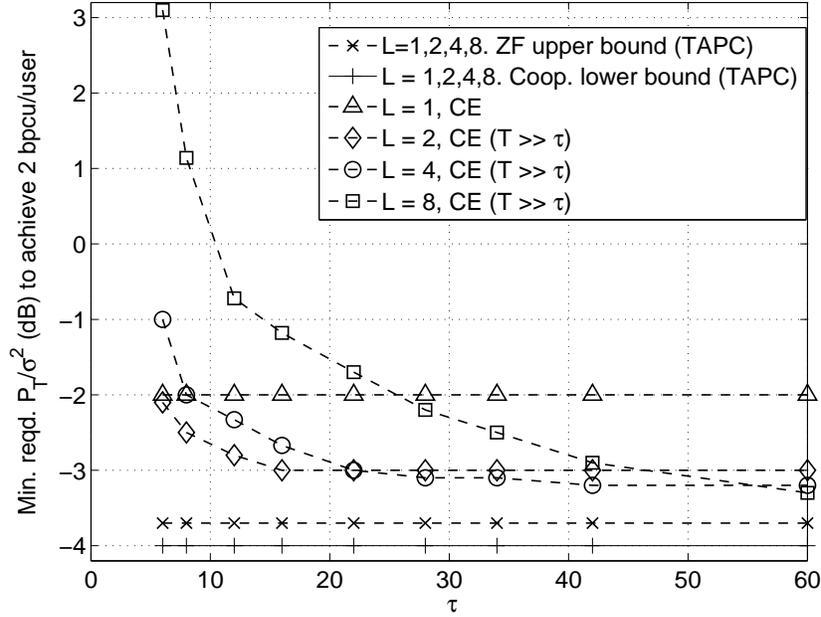, width=110mm,height=85mm}
\end{center}
\vspace{-3mm}
\caption{Minimum required $P_T/\sigma^2$ to achieve a per-user ergodic rate of $2$ bpcu, plotted as a function of $\tau$. Fixed $N=80$ BS antennas, $M=10$ users.}
\label{fig_3}
\vspace{-3mm}
\end{figure}
Fig.~\ref{fig_3} shows the minimum required $P_T/\sigma^2$ to achieve a per-user ergodic rate of $R(N,M,P_T/\sigma^2)=2$ bits-per-channel-use (bpcu),
as a function of $\tau$ ($T \gg \tau$) for different values of the channel memory $L$ and for fixed $N=80$ and $M=10$.
For the sake of comparison we also consider the performance achieved under a average only total transmit power constraint (TAPC).
Since the CE constraint is more stringent than the TAPC, it is expected
that the performance under TAPC must be better.
Under TAPC, we compute both a lower bound  (ZF - Zero Forcing precoder) and an upper bound (cooperative users) on the downlink sum capacity.
We observe that for a fixed $L$, increasing $\tau$ improves the performance.
However when $\tau \gg L$, the improvement is not so significant and the performance appears to saturate
as $\tau \rightarrow \infty$.
For $L \geq 2$, we observe that when $\tau \gg L$,  the minimum required $P_T/\sigma^2$ for the proposed CE precoding algorithm is within $1.0$ dB of the
minimum $P_T/\sigma^2$ required under TAPC. This shows the near-optimality of the proposed iterative algorithm for solving (\ref{NLS_joint_eqn}). Typically, a non-linear PA is $4-6$ times more power-efficient than a highly linear PA \cite{cripps}. Therefore using non-linear PAs with CE input signals needs $10\log_{10}(4) - 1.0 = 5.0$ dB less total transmit power compared to using highly linear PAs with a average only total transmit power constrained input. For the proposed CE precoding algorithm, it also appears that when $\tau \gg L$, the minimum required $P_T/\sigma^2$ 
is smaller for a larger $L$. 
\begin{figure}[t]
\vspace{-1mm}
\begin{center}
\epsfig{file=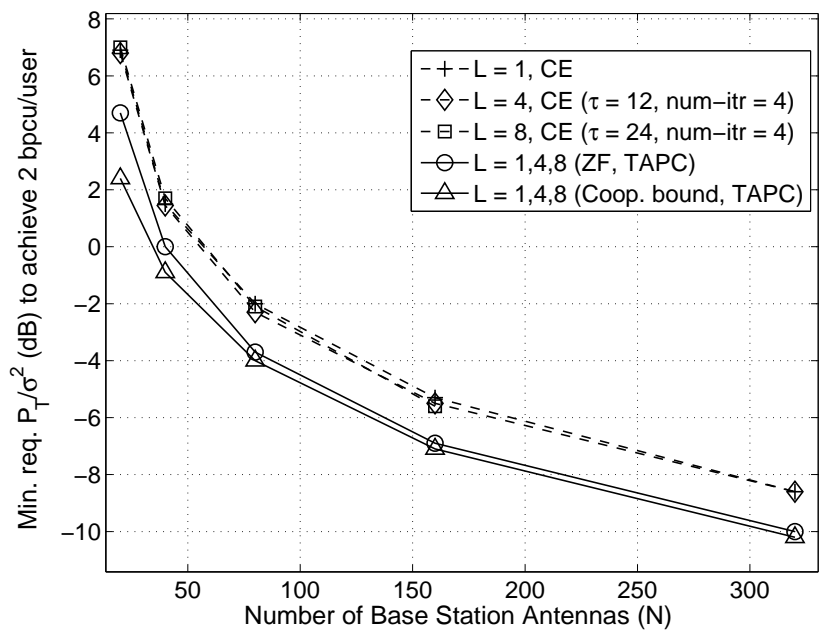, width=110mm,height=85mm}
\end{center}
\vspace{-4mm}
\caption{Minimum required $P_T/\sigma^2$ to achieve a per-user ergodic rate of $2$ bpcu, plotted as a function of $N$. Fixed $M=10$ users, $\tau = 3 L$ and $T \gg \tau$. Four iterations in the proposed CE precoding algorithm.}
\label{fig_4}
\vspace{-5mm}
\end{figure}

In Fig.~\ref{fig_4}, we plot the minimum required $P_T/\sigma^2$ to achieve $2$ bpcu/user as a function of the number of BS antennas $N$ for
a fixed number of users $M = 10$.
It can be seen that under both TAPC and CE, the minimum required $P_T/\sigma^2$ reduces by $3$ dB 
with every doubling in the number of BS antennas (holds when $N \gg M$).


\end{document}